\documentclass[aps,prfluids,reprint,superscriptaddress,onecolumn,amsmath,amssymb,nofootinbib]{revtex4-2}

\usepackage{graphicx}
\usepackage{dcolumn}
\usepackage{bm}
\usepackage{hyperref}
\hypersetup{pdftitle={Droplet mobilization} %
    pdfauthor={Sthavishtha Bhopalam, Ruben Juanes, Hector Gomez},  %
pdfproducer={lateX},
pdfview=FitV,       %
pdfstartview=FitB,
linkcolor=blue,     %
citecolor=blue,     %
urlcolor=blue,      %
breaklinks=true,    %
colorlinks=true,
citebordercolor=0 0 0,  %
filebordercolor=0 0 0,
linkbordercolor=0 0 0,
menubordercolor=0 0 0,
urlbordercolor=0 0 0,
pdfhighlight=/I,
pdfborder=0 0 0,   %
bookmarksopen=true,
bookmarksnumbered=true
}
\DeclareGraphicsExtensions{.jpg, .pdf, .tif, .png}
\graphicspath{{figures/}}%

\usepackage{lineno}  
\usepackage{xfrac}
\usepackage{soul}
\usepackage{xcolor}
\usepackage{amsfonts}
\usepackage{verbatim}
\usepackage{graphicx}
\usepackage{todonotes}
\usepackage{overpic}

\newcommand{\beq}{\begin{equation}}
\newcommand{\eeq}{\end{equation}}

\newcommand{\dt}{\partial_t}

\def\normf{\bm{n}^f}
\def\norms{\bm{n}^s}
\def\normsf{\bm{n}^{sf}}

\def\lbr{\left(}
\def\rbr{\right)}

\newcommand{\eps}{\hat{\epsilon}}

\def\velf{\bm{v}}
\def\phase{c}
\def\viscf{\eta}
\def\viscfa{\hat{\eta}_\mathrm{w}}
\def\viscfl{\hat{\eta}_\mathrm{o}}
\def\densf{\rho}
\def\densfa{\hat{\rho}_\mathrm{w}}
\def\densfl{\hat{\rho}_\mathrm{o}}

\def\fluidsf{\hat{\gamma}_\mathrm{ow}}
\def\sigmaf{\bm{\sigma}^f}

\def\dw{D_\text{w}}

\def\presfol{p^\text{ext}}
\def\presfolomg{p_\Omega}
\def\presfolomgref{\overline{p}_\Omega}
\def\wallomgref{\overline{\Omega}}
\def\prescap{\hat{p}_\text{c}}
\def\presext{\hat{p}_\text{v}}
\def\radadv{\hat{R}_\text{o,a}}
\def\radrec{\hat{R}_\text{o,r}}
\def\bodyf{\bm{H}}

\def\bodyfomgrref{\overline{H}_\omega}
\def\bodyomegaref{\overline{\omega}}
\def\bodyfr{H_r}
\def\bodyfz{H_z}
\def\timemob{t_\text{mob}}
\def\zcomd{\overline{z}_\text{d}}

\def\disps{\bm{u}}

\def\denss{\hat{\rho}^s}

\def\sigmas{\bm{\sigma}^s}
\def\pkone{\bm{P}}
\def\defgrad{\bm{F}}

\def\tubemaxrad{R_\text{m}}
\def\tubeconrad{R_\text{c}}
\def\tubeh{h}
\def\tubel{L}
\def\tubeconl{L_\text{c}}

\def\taylord{D_\mathrm{ta}}

\def\ohno{\text{Oh}}
\def\pecno{\text{Pe}}
\def\cahnno{\text{Cn}}

\def\xr{r}
\def\xz{z}
\def\Xr{R}
\def\Xz{Z}

\def\vr{v_r}
\def\vz{v_z}

\def\ur{u_R}
\def\uz{u_Z}
\def\sigmafrr{\sigma^f_{\xr \xr}}
\def\sigmafrz{\sigma^f_{\xr \xz}}
\def\sigmafzz{\sigma^f_{\xz \xz}}
\def\sigmaftt{\sigma^f_{\vartheta \vartheta}}
\def\pkonerr{P_{\Xr \Xr}}
\def\pkonerz{P_{\Xr \Xz}}
\def\pkonezz{P_{\Xz \Xz}}
\def\pkonett{P_{\vartheta \vartheta}}

\DeclareMathOperator{\tr}{tr}

\begin{document}

\title{Droplet mobilization in actuated deformable tubes}

\author{Sthavishtha R. Bhopalam}
\address{School of Mechanical Engineering, Purdue University, West Lafayette, IN 47907, USA.}
\author{Ruben Juanes}
\address{Department of Civil and Environmental Engineering, Massachusetts Institute of Technology, Cambridge, MA 02139, USA.}
\address{Department of Earth, Atmospheric and Planetary Sciences, Massachusetts Institute of Technology, Cambridge, MA 02139, USA.}
\author{Hector Gomez} \thanks{Corresponding author}
\email{hectorgomez@purdue.edu}
\affiliation{School of Mechanical Engineering, Purdue University, West Lafayette, IN 47907, USA.}

\date{\today}

\begin{abstract}

We study the mobilization of an oil droplet in a deformable, actuated constricted tube subjected to two different actuation mechanisms: hydrodynamic actuation (oscillatory body force in the fluid) and dynamic wall actuation (oscillatory traction on the tube walls). Using high-resolution fluid–structure interaction simulations, we analyze the effects of actuation frequency and amplitude on droplet transport through the constriction. Our simulations show that hydrodynamic actuation leads to a monotonic increase in the droplet’s mobilization time with increasing actuation frequency, and a decrease with increasing actuation amplitude. In contrast, dynamic wall actuation exhibits a resonance effect—the mobilization time reaches a minimum at a frequency near the tube’s resonant frequency. Our study highlights the potential of actuation mechanisms in deformable tubes for precise control of droplet transport in bio-microfluidic applications.

\end{abstract}

\maketitle

\section{Introduction}

The motion of droplets and bubbles in confined geometries has long attracted fundamental and practical interest. This problem has been primarily studied by researchers interested in dislodging oil ganglia trapped in narrow pore throats by seismic stimulation \cite{beresnev_johnson_geophys_1994}---an actuation technique that applies low–frequency elastic waves to mobilize residual oil over long distances \cite{white_undergroundsound_2000}. This actuation can be represented as an oscillatory body force superposed on a steady pressure gradient inside a capillary tube with a constriction that is defined by a smooth variation of the tube's cross-section \cite{beresnev_geophys_2006, beresnev_deng_seg_2010}. Studies based on this configuration show that oil droplets in the tube can be mobilized beyond the constriction when the tube is actuated at sufficiently low frequencies and/or high amplitudes \cite{li_etal_jcis_2005, graham_higdon_jfm_2000}.

Although droplet mobilization has been previously studied in actuated rigid constrictions, the corresponding problem in an actuated deformable constriction has received less attention. A fundamental study of droplet mobilization in an actuated deformable constriction is important for three reasons: (a) the tube’s deformability provides an additional control parameter for the droplet mobilization process; (b) the tube’s deformability enables new actuation strategies, for example, surface-acoustic-wave (SAW) forcing induced vibrations and deformations of the capillary tube, that can have a large impact on the droplet transport dynamics \cite{wang_etal_acsappl_2021, zhou_aprev_2023}; and (c) droplet or bubble motion through deformable narrow tubes is common in many biological and engineering applications \cite{grotberg_arfm_1994}, including arterial vasomotion \cite{murdock_etal_nat_2024}, air embolism in microvessels \cite{li_etal_pnas_2021}, cardiopulmonary bypass \cite{hugenroth_etal_scirep_2021}, plant leaf vessels \cite{keiser_etal_jfm_2022, gauci_etal_intfoc_2025}, enhanced-oil-recovery processes \cite{beresnev_johnson_geophys_1994, huh_spe_2006}, drug-delivery systems \cite{abundo_etal_pnas_2025}, and soft robotics \cite{matia_advintsys_2023}.

Existing approaches for understanding droplet mobilization rely primarily on experiments \cite{li_etal_jcis_2005, zhang_etal_jgr_2019} and quasi-steady theoretical models with simplifications \cite{deng_cardenas_wrr_2013, beresnev_geophys_2006}. Experiments are challenging because they are time-consuming, need advanced microfabrication setup and real-time feedback control. Theoretical models, however, rely on several assumptions: they typically assume spherical shape of the fluid-fluid interfaces, they neglect dynamic contact-angle effects, some impose fully developed flow approximations that may be valid only at high actuation frequencies, while some assume that the droplet is completely lubricated by a thin film of another immiscible fluid \cite{beresnev_geophys_2006}. As shown in \cite{zhang_etal_jgr_2019}, such assumptions can lead to large deviations from experiments. These limitations motivate the use of rigorously validated high-resolution computational models without the aforementioned assumptions.

To address this gap, we simulate the displacement of an oil droplet by water in an actuated, deformable capillary tube with a smoothly varying constriction, using a multicomponent fluid–structure-interaction model. In the present work, we study two actuation mechanisms: hydrodynamic actuation and dynamic wall actuation. In hydrodynamic actuation, we prescribe an oscillatory body force in the fluid, similar to the previous studies in a rigid tube. However, in dynamic-wall actuation, we impose an oscillatory external pressure on the tube walls---this actuation setup can be experimentally realized using piezoelectric-generated tube vibrations or by using acoustic waves. {Our computations show that, under hydrodynamic actuation, the droplet mobilization time increases monotonically with actuation frequency and decreases monotonically with the actuation amplitude}. Under dynamic-wall actuation, however, the mobilization time is minimum within a narrow range of actuation frequencies. Under actuation within this frequency range, the system displays features of resonance and the droplet mobilization is fastest. The mobilization time decreases progressively as the deviation from the resonant actuation frequency increases. Additionally, in the dynamic wall actuation case, the droplet mobilization time decreases monotonically with the actuation amplitude. Our findings highlight the importance of controlling actuation parameters in optimizing droplet transport through constricted capillary tubes. These findings are important in many applications, from including enhancing oil recovery in subsurface flows to designing microfluidic systems and SAW-driven platforms for on-demand droplet manipulation, mixing and release.

\section{Model of droplet mobilization}

\subsection*{Mechanism of droplet mobilization in a constricted capillary tube}

\begin{figure}[t] 
    \centering
    \includegraphics[width=\linewidth]{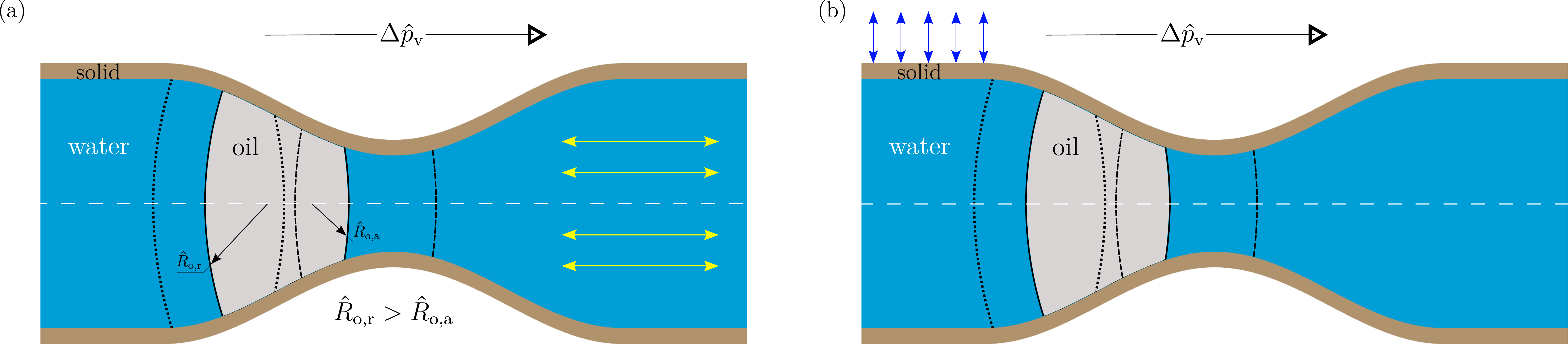}
    \caption{Mobilization of a droplet in a constricted deformable capillary tube subjected to two distinct oscillatory actuation mechanisms. Panel (a) shows the hydrodynamic actuation mechanism via oscillatory forcing in the fluid (arrows shown in yellow). Panel (b) shows the dynamic wall actuation mechanism enabled by imposing a follower load on the tube wall (arrows shown in blue). In panels (a–b), the droplet’s initial position is shown in gray. The black dashed lines indicate the droplet’s position when the inertial component from the actuation is in the direction of the external pressure drop $\Delta \presext$, whereas the black dotted lines denote its position when the inertial component from the actuation opposes $\Delta \presext$. The white dotted line represents the centerline axis of the tube. The droplet size in the figure at different positions is not drawn to scale.}
    \label{fig:concept}
\end{figure}

Consider an oil droplet in a constricted capillary tube filled with a less viscous immiscible fluid such as water; see Fig.~\ref{fig:concept}. A capillary resistance is established across the constriction. If we assume that the fluid-fluid interfaces are spherical caps, the resistance can be quantified by the capillary pressure drop $\Delta \prescap$. Here, and in what follows, we use the superscript $\hat{}$ to denote dimensional variables. We omit that superscript to indicate the corresponding dimensionless
variables. Then, we can write
\begin{equation}
    \Delta \prescap = 2 \fluidsf \lbr \frac{1}{\radadv \lbr \theta_\mathrm{a} \rbr} - \frac{1}{\radrec \lbr \theta_\mathrm{r} \rbr} \rbr,
    \label{eqn:capillary_presdrop}
\end{equation}
$\!\!\!$ where $\fluidsf$ is the surface tension at the oil-water interface, $\radadv$ and $\radrec$ are the principal radii of curvature of the advancing and receding droplet-water menisci, and $\theta_\mathrm{a}$ and $\theta_\mathrm{r}$ are the advancing and receding contact angles made by the droplet-water interface with the tube walls. An external pressure drop, denoted by $\Delta \presext$, is imposed across the tube ends to drive the droplet. To illustrate the problem, we make the following assumptions in this section: (a) quasi-static equilibrium of the droplet-water interface, (b) negligible contact angle hysteresis, i.e., $\theta_\mathrm{a} = \theta_\mathrm{r}$, and (c) negligible viscous pressure losses across the droplet. 

When the droplet is in the straight section of the tube, the curvatures of the advancing and receding droplet-water menisci are equal. As a result, $\Delta \prescap = 0$, and the droplet motion is governed solely by the imposed pressure drop. However, when the droplet enters the converging section of the constriction, the curvature of the receding meniscus becomes smaller than that of the advancing meniscus; see Fig.~\ref{fig:concept}. Depending on the balance between the imposed pressure drop and the capillary resistance, two outcomes are possible \cite{deng_cardenas_wrr_2013, beresnev_grl_2005}: (a) droplet trapping, when $\Delta \prescap > \Delta \presext$, i.e., the capillary pressure drop exceeds the driving force, and the droplet remains immobile; and (b) droplet mobilization, when $\Delta \prescap < \Delta \presext$, i.e., the driving force overcomes the capillary resistance, and the droplet advances up to the neck of the constriction---this phenomenon was first observed by Jamin \cite{jamin_1860}. Here, we define the neck of the constriction as the point where the tube's cross-sectional area is the smallest. Once the droplet passes into the diverging section of the constriction, the capillary pressure difference becomes negative, i.e., $\Delta \prescap < 0$. The capillary pressure drop now assists the droplet motion, so there is no further resistance to droplet advancement beyond the neck of the constriction. From Eq.~\eqref{eqn:capillary_presdrop}, we conclude that, in addition to $\fluidsf$, the entrapment of the droplet in the constriction is also affected by: (a) {wettability---wetting droplets are easier to mobilize than non-wetting droplets in the converging section of a rigid constriction}; and (b) geometry of the constriction---narrower constrictions can trap droplets more effectively due to the larger capillary pressure drop.

Previous studies on droplet dynamics in a constricted rigid capillary tube \cite{beresnev_grl_2005, beresnev_geophys_2006, graham_higdon_jfm_2000, li_etal_jcis_2005, deng_cardenas_wrr_2013} have shown that axial vibrations of the tube can overcome capillary resistance and cause droplet mobilization under droplet trapping conditions. This phenomenon can be explained as follows: Axial vibrations of the tube induce an inertial forcing on the fluid. During each oscillatory cycle, when this inertial force aligns with the imposed pressure drop, the droplet moves forward. Conversely, when the inertial force opposes the imposed pressure drop and overpowers the pressure drop, the droplet moves backward. After several oscillatory cycles, the net forward displacement of the droplet accumulates, eventually driving the droplet past the tube’s constriction. In the present work, we propose two actuation mechanisms to mobilize droplets in a deformable constricted capillary tube: (a) hydrodynamic actuation, where an oscillatory body force is applied to the fluid [Fig.~\ref{fig:concept}(a)]; and (b) dynamic wall actuation, where a follower-load traction is imposed on the tube walls [Fig.~\ref{fig:concept}(b)]. These two mechanisms modulate the local pressure distribution in the fluid, thereby controlling the droplet’s advancing or receding motion. {For example, in the case of dynamic wall actuation, when the applied wall traction causes the tube to contract, the droplet advances forward. Conversely, when the traction causes the tube to expand, the droplet moves backward.} To study the effect of these two actuation mechanisms on droplet motion in a deformable constricted capillary tube, we use a multi-component fluid-structure interaction (FSI) model \cite{bhopalam_cmame_2022, bhopalam_imbibdrainage_2025, bueno_2018b}. {Our simulations show that the dynamic wall actuation exhibits a resonance-like response, where droplet mobilization is the fastest at specific actuation frequencies due to the coupled interactions between the fluid and tube.} To reduce the computational cost of our simulations, we simplify our three-dimensional FSI model by assuming axisymmetry. We use the Navier-Stokes-Cahn-Hilliard equations \cite{anderson_arfm_1998} to model the mixture of oil and water, by replacing the oil-water interface with a thin diffuse region. We model the solid as an isotropic hyperelastic material. 

\subsection*{Key dimensionless numbers governing droplet mobilization}

To study the droplet mobilization problem in a deformable capillary tube, we non-dimensionalize the variables with reference length $\hat{L}_\mathrm{r}$, reference time $\hat{t}_\mathrm{r} = \frac{\viscfl \hat{L}_\mathrm{r}}{\fluidsf}$ and reference mass $\hat{m}_\mathrm{r} = \densfl \hat{L}_\mathrm{r}^3$, respectively. Here, $\viscfl$ is the dynamic viscosity of oil and $\densfl$ is the true density of oil (i.e., mass of oil over the volume occupied by oil), respectively. We classify our dimensionless numbers into four different types based on whether they relate to geometric, transport, dynamic, or material properties. We define the following geometric dimensionless numbers: (a) radial thickness of the tube $h$, (b) axial length of the tube $L$, (c) axial length of the tube's constriction $L_\mathrm{c}$, (c) tube's initial radius at the neck of the constriction $R_\mathrm{c}$, and (d) Cahn number $\cahnno = \frac{\eps}{\hat{L}_\mathrm{r}}$, where $\eps$ is the diffuse interface length scale. Our transport dimensionless number is the P\'eclet number $\pecno = \frac{\hat{L}_\mathrm{r}^2}{\hat{M} \viscfl}$ defined as the ratio of advection to diffusion. Here, $\hat{M}$ is the positive mobility coefficient that enters the Navier-Stokes-Cahn-Hilliard equations. We define the following dynamic dimensionless variables for the fluid and the solid: (a) Ohnesorge number $\ohno = \frac{\viscfl}{\sqrt{\densfl \fluidsf \hat{L}_\mathrm{r}}}$ which is the ratio of inertio-capillary to the inertio-viscous time scale; (b) viscosity ratio of oil to water $\tilde{\viscf} = \frac{\viscfl}{\viscfa}$, where $\viscfa$ is the dynamic viscosity of water; (c) density ratio of oil to water $\tilde{\densf} = \frac{\densfl}{\densfa}$, where $\densfa$ is the true density of water; (d) elastocapillary number $\chi = \frac{\fluidsf}{\hat{E} \hat{L}_\mathrm{r}}$ which quantifies the strength of elastocapillary effects, where $\hat{E}$ is the Young's modulus of the solid; (e) bendocapillary number \cite{roman_bico_jp_2010, style_arcmp_2017} given by $\zeta = \frac{\hat{E} \hat{h}^3}{\fluidsf \hat{L}_\mathrm{r}^2}$. Our dimensionless number defining the tube's material property is the Poisson's ratio $\nu$.

\subsection{Governing equations of fluid mechanics}
\label{sec:fluidmech}

We use the Navier-Stokes-Cahn-Hilliard (NSCH) equations, a phase-field model \cite{abels_etal_2012, gomez_etal_frontiers_2023, khanwale_cpc_2022} that describes the dynamics of the two immiscible fluids: oil and water. We assume that oil and water are individually incompressible, and share the same velocity field. We neglect gravitational forces. The dimensionless governing equations of the fluid mixture in Eulerian coordinates are given by
\begin{subequations}
    \begin{alignat}{2}
        & \frac{1}{\xr} \frac{\partial}{\partial \xr} (\xr \vr) + \frac{\partial \vz}{\partial \xz} = 0 \label{eqn:continuity_eqn}, \\
        &\begin{aligned}
            \rho &\left(\dt \vr + \vr \frac{\partial \vr}{\partial \xr} + \vz \frac{\partial \vr}{\partial \xz} \right) 
            = \frac{\partial \sigmafrr}{\partial \xr}
            + \frac{1}{\xr} \lbr \sigmafrr - \sigmaftt \rbr + \frac{\partial \sigmafrz}{\partial \xz} + \bodyfr, \\ 
        \end{aligned}\label{eqn:momentum_eqn_r} \\
        &\begin{aligned}
            \rho &\left(\dt \vz + \vr \frac{\partial \vz}{\partial \xr} + \vz \frac{\partial \vz}{\partial \xz} \right) 
            = \frac{\partial \sigmafrz}{\partial \xr} + \frac{1}{\xr}\sigmafrz + \frac{\partial \sigmafzz}{\partial \xz} + \bodyfz, \\ 
        \end{aligned} \label{eqn:momentum_eqn_z} \\
        & \dt \phase + \vr \frac{\partial \phase}{\partial \xr} + \vz \frac{\partial \phase}{\partial \xz} = \frac{1}{\pecno} \Bigg( \frac{1}{\xr} \frac{\partial}{\partial \xr} \lbr \xr \frac{\partial \mu}{\partial \xr} \rbr  + \frac{\partial^2 \mu}{\partial \xz^2} \Bigg), \\
        & \mu = \frac{3}{2 \sqrt{2}} \Bigg[ \frac{1}{\cahnno} \phase \lbr \phase^2 - 1 \rbr - \cahnno \Bigg( \frac{1}{\xr} \frac{\partial}{\partial \xr} \lbr \xr \frac{\partial \phase}{\partial \xr} \rbr + \frac{\partial^2 \phase}{\partial \xz^2} \Bigg) \Bigg],
        \label{eqn:chempot_eqn} 
    \end{alignat}
    \label{eqn:fluidmech}
\end{subequations}
$\!\!\!$where $\xr$ is the radial spatial coordinate, $\xz$ is the axial spatial coordinate, $\velf = \lbr \vr, \vz \rbr$ is the fluid velocity where the subscripts $\xr$ and $\xz$ denote the radial and axial directions of the spatial domain, $\densf$ is the density of the fluid mixture, $\dt$ denotes partial time differentiation, $\sigmaf = \left[\sigma^f_{ij}\right]_{i,j=\xr, \vartheta, \xz}$ is the fluid Cauchy stress tensor where $\vartheta$ denotes the azimuthal direction, $\bodyf = \lbr \bodyfr, \bodyfz \rbr$ is the body force in the fluid, $\phase \in \left[-1, 1\right]$ is the phase field, and $\mu$ is the chemical potential. {In what follows, we use $\phase = -1$ to denote water and $\phase = 1$ to denote oil}. In Eqs.~\eqref{eqn:momentum_eqn_r} and ~\eqref{eqn:momentum_eqn_z}, 
$\densf = \frac{1}{2} \lbr 1 + \frac{1}{\tilde{\densf}}\rbr + \frac{1}{2} \lbr 1 - \frac{1}{\tilde{\densf}}\rbr \phase$ and 
\begin{equation*}
    \sigmaf = - p \bm{I} + 2 \ohno^2 \viscf(\phase)  \ \nabla^{s} \bm{v} - \frac{3}{2\sqrt{2}} \ \cahnno \ \ohno^2 \ \nabla c \otimes \nabla c,
\end{equation*}
where $p$ is the fluid pressure, $\nabla^{s}$ is the symmetrized form of the spatial gradient operator $\nabla$, and $\viscf (\phase) = \frac{1}{2}\lbr 1 + \frac{1}{\tilde{\viscf}} \rbr + \frac{1}{2}\lbr 1 - \frac{1}{\tilde{\viscf}} \rbr c$ is the mixture viscosity. In the case of {hydrodynamic actuation}, we use $\bodyfr = 0$ and $\bodyfz = \bodyfomgrref \sin \lbr 2 \pi \bodyomegaref t \rbr$, where $\bodyfomgrref = \frac{\hat{h}_\omega \hat{R}_\mathrm{m}^2}{\hat{\gamma}_\mathrm{ow}}$ is the dimensionless forcing amplitude and $\bodyomegaref = \frac{\viscfl \hat{R}_\mathrm{m} \hat{\omega}}{\hat{\gamma}_\mathrm{ow}}$ is the dimensionless forcing frequency. Here, we have non-dimensionalized $\hat{\omega}$ using the viscous response time of the droplet $\frac{\viscfl \hat{R}_\mathrm{m}}{\hat{\gamma}_\mathrm{ow}}$, where we assume that the initial droplet size is on the order of $\hat{R}_\mathrm{m}$. However, in the case of dynamic wall actuation, we select $\bodyf = \bm{0}$.

\subsection{Governing equations of solid mechanics}

We describe the solid dynamics using the dimensionless linear momentum balance equation written in the reference coordinates, 
\begin{subequations}
    \begin{alignat}{2}
        & \left. \dt\dt \ur \right|_{\bm{X}} = \frac{\partial \pkonerr}{\partial \Xr} + \frac{1}{\Xr} \lbr \pkonerr - \pkonett \rbr + \frac{\partial \pkonerz}{\partial \Xz},
        \label{eqn:momentum_eqn_solid_1}
        \\
        & \left. \dt\dt \uz \right|_{\bm{X}} = \frac{\partial \pkonerz}{\partial \Xr} + \frac{1}{\Xr}\pkonerz + \frac{\partial \pkonezz}{\partial \Xz},
        \label{eqn:momentum_eqn_solid_2}     
    \end{alignat}
    \label{eq:solidmech}
\end{subequations}
$\!\!\!$where we use the spatial derivative with respect to the radial and axial reference coordinates $\Xr$ and $\Xz$, $\disps = \lbr \ur, \uz \rbr$ is the solid displacement and $\pkone = \left[P_{ij} \right]_{i,j=\Xr,\vartheta,\Xz}$ is the first Piola-Kirchhoff stress tensor. We compute the second-order time derivative in Eq.~\eqref{eq:solidmech} by holding $\bm{X} = \lbr \Xr, \Xz \rbr$ fixed. We model the solid as a nonlinear, hyperelastic, isotropic and homogeneous material. To describe the material behavior of the solid, we use a neo-Hookean constitutive model \cite{simo_2006_book}. The strain energy per unit volume of the solid in the reference configuration is given by 
\begin{equation*}
    \hat{W} = \frac{\hat{\kappa}}{2} \left(\frac{1}{2}\left(J^2 - 1\right) - \ln{J}\right) + \frac{\hat{G}}{2}\left(J^{-\sfrac{2}{3}}\tr\left({\bm{C}}\right) - 3 \right),
\end{equation*}
where $\hat{\kappa}$ and $\hat{G}$ are the bulk and shear moduli of the solid and $\bm{C} = \defgrad^T\defgrad$ is the right Cauchy-Green deformation tensor. The Young's modulus and Poisson ratio are defined from $\hat{\kappa}$ and $\hat{G}$ as $\hat{E} = \frac{9 \hat{\kappa} \hat{G}}{3 \hat{\kappa} + \hat{G}}$ and $\nu = \frac{3\hat{\kappa} - 2 \hat{G}}{2 \left(3 \hat{\kappa} + \hat{G} \right)}$. In Eq.~\eqref{eq:solidmech}, we compute $\pkone$ from 
\begin{equation}
\pkone = 2 \defgrad \frac{\partial W}{\partial \bm{C}}= 
\frac{\ohno^2}{3 \chi \lbr 1 - 2\nu \rbr} \frac{\densfl}{\denss} \bm{\defgrad} J^{-\sfrac{2}{3}}\Bigl(\bm{I} - \frac{1}{3} \tr\lbr {\bm{C}} \rbr \bm{C}^{-1}\Bigr) + \frac{\ohno^2}{6 \chi \lbr 1 + \nu \rbr} \frac{\densfl}{\denss} \bm{\defgrad} \Big( J^{2} - 1 \Big) \bm{C}^{-1},
\end{equation}
where $\denss$ is the mass density of the solid in the reference configuration.

\subsection{Initial and boundary conditions of the coupled fluid-structure interaction problem}

\begin{figure}[t] 
    \centering
    \includegraphics[width=0.9\linewidth]{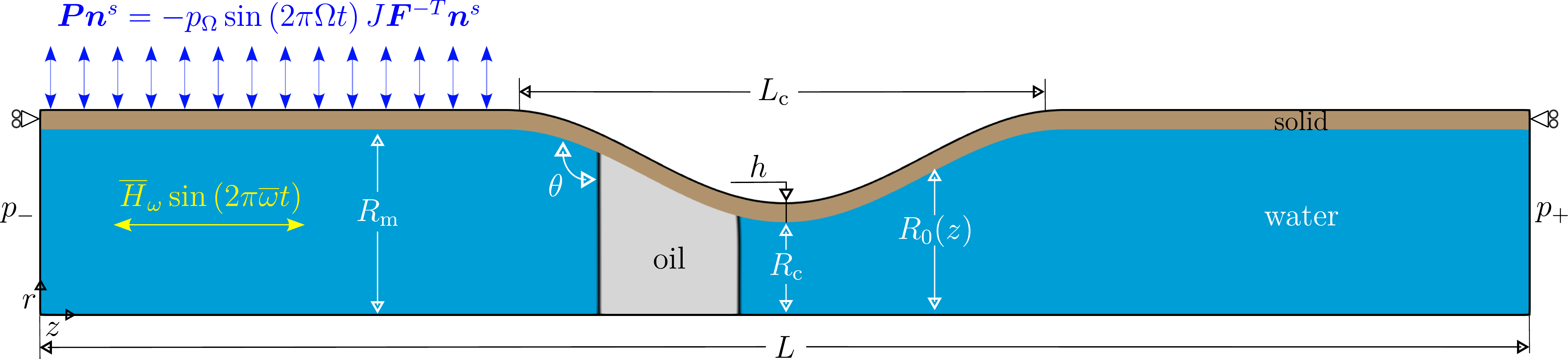}
    \caption{Schematic of the two-dimensional axisymmetric computational domain, initial conditions and geometrical parameters used in our numerical simulations. The oscillatory fluid-forcing case is enabled by applying an oscillatory body force in the fluid (yellow arrows). The dynamic wall-actuation case is enabled by imposing a follower-load boundary condition on the tube wall (blue arrows).}
    \label{fig:schematic}
\end{figure}

Fig.~\ref{fig:schematic} depicts our axisymmetric computational domain and the tube's geometry, which we approximate with a smoothly varying constriction such that the tube’s initial radius is defined as
\begin{equation}
    R_0 (z) = \begin{cases}
        \tubemaxrad & \text{if} \ \xz \leq \frac{1}{2}\lbr \tubel - \tubeconl \rbr \ \text{and} \ \xz \geq \frac{1}{2}\lbr \tubel + \tubeconl \rbr, \\
        \frac{\tubeconrad}{2} \Bigg( \lbr 1 + \frac{\tubemaxrad}{\tubeconrad} \rbr + \lbr 1 - \frac{\tubemaxrad}{\tubeconrad} \rbr \cos \Big( \frac{2 \pi}{\tubeconl} \xz - \frac{\pi} \tubeconl \tubel \Big) \Bigg) & \text{if} \ \frac{1}{2} \lbr \tubel - \tubeconl \rbr < \xz < \frac{1}{2} \lbr \tubel + \tubeconl \rbr.
    \end{cases}
    \label{eqn:tuberad}    
\end{equation}
Here, $\tubemaxrad$ is the maximum initial radius of the capillary tube. Our tube's geometry is motivated by the early studies on droplet mobilization through rigid constricted tubes \cite{beresnev_geophys_2006, deng_cardenas_wrr_2013}. In our numerical simulations, we specify the fluid velocity $\velf_0$ and the phase-field $\phase_0$ at the initial time for the fluid mechanics equations presented in Eq.~\eqref{eqn:fluidmech}. For the solid mechanics equations [see Eq.~\eqref{eq:solidmech}], we prescribe the initial solid displacement $\disps_0$ and initial solid velocity. 

We now discuss the boundary conditions of our problem. The oscillatory actuation mechanism in our simulations can cause periodic flow reversal along the lateral fluid boundaries. Imposing the natural outflow (Neumann) conditions on these boundaries will then generate spurious wave reflections and induce numerical instabilities that allow unphysical energy influx across these boundaries. To maintain numerical stability in the presence of flow reversal, we apply a backflow stabilized boundary condition of the form \cite{moghadam_compmech_2011, neighbor_ewco_2023}
\begin{equation}
    \sigmaf \normf = 
    - p_{\pm} \normf + 2 \ohno^2 \viscf(\phase) \ \nabla^{s} \bm{v} \normf + \beta \densf \lbr \velf \cdot \normf \rbr^- \velf,     
    \label{eqn:backflow_bc}
\end{equation}
where $\lbr \velf \cdot \normf \rbr^- = \frac{\velf \cdot \normf - |\velf \cdot \normf|}{2}$, $\normf$ is the unit outward normal vector to the external fluid boundary, $p_{\pm}$ denotes the imposed fluid pressure at the left ($p_-$) and right ($p_+$) lateral walls of the capillary tube and $\beta$ is a stabilization term. 
This stabilized boundary condition was proposed in \cite{bazilevs_cmame_2009}, and has since been successfully used in multiple applications including blood‐flow simulations \cite{neighbor_ewco_2023}. Eq.~\eqref{eqn:backflow_bc} combines pressure traction and viscous traction contributions, with an additional advective traction term that opposes backflow on the lateral fluid boundaries. Although this advective traction term can slightly change the pressure at these boundaries, our numerical simulations confirm that its impact on the overall solution is negligible. In Eq.~\eqref{eqn:backflow_bc}, we have neglected the capillary traction contributions since their influence near the lateral fluid boundaries is negligible. Unless otherwise specified, we set $\beta = 0.01$ in Eq.~\eqref{eqn:backflow_bc}. On the lateral fluid boundaries, we additionally enforce neutral wettability $\normf \cdot \nabla \phase = 0$, and zero diffusive flux $\normf \cdot \nabla \mu = 0$. {At the bottom fluid boundary, we impose a no penetration condition for the fluid velocity $\velf \cdot \normf = 0$, along with $\normf \cdot \nabla \phase = 0$, $\normf \cdot \nabla \mu = 0$ and zero axial traction $\bm{e}_z \cdot \sigmaf \normf = 0$. Here, $\bm{e}_z$ is a unit vector along the axial direction. On the lateral solid boundaries, we impose a zero axial displacement $\uz = 0$ and a radial traction-free condition $\bm{e}_R \cdot \pkone \norms = 0$, where $\norms$ is the unit outward normal vector to the external solid boundary and $\bm{e}_R$ is a unit vector along the radial direction}. At the top solid boundary, we set the traction-free condition $\pkone \norms = 0$ for the case of {hydrodynamic actuation}. However, for dynamic wall actuation case, we prescribe a follower load boundary condition only to a portion of the capillary tube; see Fig.~\ref{fig:schematic}. This boundary condition is of the form $\pkone \norms = - \presfol J \defgrad^{-T} \norms$ where 
\begin{equation}
\presfol = 
    \begin{cases}
        {\frac{\presfolomgref \tubeh E}{R_\mathrm{m} + \tubeh} \sin \lbr 2 \pi \Omega t \rbr} & \text{if} \ \xz \leq \frac{1}{2}\lbr \tubel - \tubeconl \rbr, \\
        0 & \text{if} \ \xz > \frac{1}{2} \lbr \tubel - \tubeconl \rbr. 
    \end{cases}
    \label{eqn:followerloadpres}
\end{equation}
{Here, $\presfolomgref = \frac{\hat{p}_\Omega \lbr \hat{R}_\mathrm{m} + \hat{\tubeh} \rbr}{\hat{\tubeh} \hat{E}}$ is the traction amplitude ratio and $\Omega$ is the traction frequency. We define the traction frequency ratio $\wallomgref = \frac{\hat{\Omega}}{\hat{\Omega}_\text{res}}$, where $\hat{\Omega}_\text{res} = \frac{\hat{\tubeh}}{\lbr \hat{R}_\mathrm{m} + \hat{\tubeh} \rbr^2} \sqrt{\frac{\hat{E}}{12 \denss \lbr 1 - \nu^2 \rbr}}$ is the natural frequency of an annular tube \cite{soedel_book_2004}, which we have obtained under the assumption of a long cylindrical tube with no external loading and only considering the leading-order terms}. The choice of our pressure scale $\frac{\hat{\tubeh} \hat{E}}{\lbr \hat{R}_\mathrm{m} + \hat{\tubeh} \rbr}$ for non-dimensionalizing $\hat{p}_\Omega$ in the definition of $\presfolomgref$ is detailed in Appendix \ref{sec:appendix_a}. On the fluid-solid interface, we enforce the following coupling conditions: (a) kinematic compatibility of the fluid and solid velocities $\velf = \dt \disps$; (b) balance of the fluid and solid tractions $\sigmaf \normsf = \sigmas \normsf$, where $\normsf$ is the unit normal vector to the fluid-solid interface in the direction from fluid to solid and $\sigmas$ is the solid Cauchy stress tensor computed as $\sigmas = J^{-1} \pkone \defgrad^{T}$; (c) zero diffusive flux; and the (d) dynamic wettability condition \cite{yue_feng_pof_2011}
\begin{equation}
    \normsf \cdot \nabla \phase = \frac{2 \sqrt{2}}{3\cahnno} \lbr - \dw \big(\dt \phase + \velf \cdot \nabla \phase \big) + \frac{3}{4} \lbr 1 - \phase^2 \rbr \cos \theta \rbr.
    \label{eqn:dynamic_contactangle_law}
\end{equation}
Here, $\dw$ is a non-negative dynamic wall mobility coefficient and $\theta$ is the equilibrium contact angle made by the oil-water interface with the solid, which we assume to be a constant; see Fig.~\ref{fig:schematic}. A non-zero value of the dynamic wall mobility coefficient $\dw$ captures deviations of the advancing and receding contact angles from the equilibrium angle $\theta$.

\subsection{Computational method}

We solve the governing equations using a body-fitted fluid-structure algorithm, similar to the approach used in \cite{bueno_2018b, bhopalam_cmame_2022}. First, we rewrite the fluid governing equations given by Eq.~\eqref{eqn:fluidmech} and the boundary conditions in Arbitrary Lagrangian-Eulerian form \cite{donea_2004}. We then recast the governing equations of the FSI problem in weak form and spatially discretize them using Isogeometric Analysis \cite{cottrell_wiley_2009, hughes_cmame_2005}. We use a space of splines that is conforming at the fluid-solid interface. We perform time integration using the generalized-$\alpha$ method \cite{jansen_2000}. {We use a time-step controller that dynamically adjusts the time step according to the number of nonlinear solver iterations required for convergence.} To ensure good resolution of the oscillatory actuation, we set the maximum allowed time step $\Delta \hat{t}$ to $\frac{1}{100\hat{\omega}}$ for fluid forcing actuation and $\frac{1}{100\hat{\Omega}}$ for dynamic wall actuation. We develop our code on top of the open source high-performance computing libraries PETSc \cite{petsc-web-page} and PetIGA \cite{PETiga_CMAME}. We have validated our computational method on benchmark fluid–solid interaction problems, such as, capillary folding of elastic sheets by droplets \cite{bueno_2018b, bhopalam_cmame_2022}, static wetting experiments of droplets on silicone gel \cite{bueno_2018b}, spontaneous droplet motion on soft solids \cite{bhopalam_softmat_2024}, and forced imbibition and drainage in capillary tubes \cite{bhopalam_imbibdrainage_2025}. 

\section{Results}

Using high-resolution numerical simulations, we investigate the mobilization of an oil droplet in a deformable, constricted capillary tube under hydrodynamic actuation and dynamic wall actuation; see Fig.~\ref{fig:schematic}. We choose a reference length scale of $\hat{L}_\mathrm{r} = 375 \ \mu$m to match the droplet microfluidics experiments in a rigid capillary tube reported in \cite{zhao_etal_prl_2018, pahlavan_etal_pnas_2019}. Unless otherwise specified, we define the tube geometry by specifying five non-dimensional parameters: $\tubemaxrad = 1$, $\tubeh = 0.1$, $\tubel = 8$, $\tubeconl = 3$, and $\tubeconrad = 0.5$. We select $\tubel$ to be large enough compared to $\tubeconl$ to reduce the boundary effects on the droplet motion. We selected these geometric parameters to match the ranges employed in previous studies of droplet motion through rigid constricted tubes \cite{beresnev_geophys_2006} while also ensuring that our simulations remain computationally affordable.

We select the fluid properties corresponding to an oil droplet in water \cite{deng_cardenas_wrr_2013} with $\viscfa = 10^{-3} \ \mathrm{Pa} \cdot \mathrm{s}$, $\densfa = 1000 \ \mathrm{kg} \cdot \mathrm{m^{-3}}$, oil-water surface tension of $50 \ \mathrm{mN/m}$, $\tilde{\viscf} = 10$ and $\tilde{\densf} = 0.95$. We prescribe the wettability of the oil droplet by selecting $\theta = 30^\circ$ (non-wetting) and $\dw = 0.75$. The value of $\dw$ is adopted from our previous study \cite{bhopalam_imbibdrainage_2025}, where we showed that it reproduces the contact-line speeds observed in capillary tube experiments \cite{zhao_etal_prl_2018}. {We selected a non-wetting droplet here because the Laplace pressure remains positive, thereby inducing tube expansion in the traction-free wall case and promoting droplet advancement through the constriction. Had we selected a wetting droplet, the tube and the neck of the constriction might have contracted in the traction-free wall case, potentially impeding droplet advancement towards the constriction; for more details, see the discussion in \cite{bhopalam_imbibdrainage_2025} for a soft capillary tube and \cite{bradley_etal_prl_2019} for a deformable channel}. For our selected fluid properties, we have $\ohno = 0.073$ and $\pecno = 2814$. We choose $\cahnno = 0.015$ to ensure that the diffuse oil-water interface is sufficiently thin and the criterion for the sharp-interface limit $\cahnno < 4 \pecno^{-0.5} \tilde{\viscf}^{-0.25}$ is satisfied \cite{yue_etal_jfm_2010}. For the deformable tube, we select properties typical of PDMS elastomers \cite{palchesko_plos_2012}: $\denss = 1000 \ \mathrm{kg\, m^{-3}}$, $\nu = 0.45$ and $\hat{E} = 100$ $\text{kPa}$. For our selected fluid and solid properties, we get $\chi = 1.33 \times 10^{-3}$ and $\zeta = 0.75$. 

In all simulations, we use a uniform mesh with $\tubel \times (\tubemaxrad + \tubeh) \times 10^4$ quadratic elements. Our initial conditions in the simulation are as follows: $\disps_0 = {0}$, $\velf_0 = 0$ and we assume that the oil-water interface is initially flat, such that $\phase_0 (z)= 1 - \tanh \frac{\lbr \xz – z_\text{in,l}\rbr}{\sqrt{2} \epsilon} + \tanh \frac{\lbr \xz – z_\text{in,r}\rbr}{\sqrt{2} \epsilon}$. Here, $z_\text{in,l}$ and $z_\text{in,r}$ are the receding and advancing axial positions of the initial oil-water interface, respectively. We select $z_\text{in,l} = 3$ and $z_\text{in,r} = 3.75$; see Fig.~\ref{fig:schematic}. To trigger droplet mobilization in our simulations, we impose an external pressure drop with $p_- = 0.016$ at the inlet and $p_+ = 0$ at the outlet of the capillary tube. Under these conditions, and without any actuation mechanisms, the droplet can overcome the capillary resistance and move through the constriction. We track two different droplet transport regimes: (a) partial droplet mobilization, in which the droplet breaks up before it crosses the constriction's neck and (b) complete droplet mobilization, in which the droplet remains intact. In the case of complete droplet mobilization, we define the mobilization time, $\timemob$, as the time interval required for the droplet’s axial center of mass $\zcomd$ to cross the constriction, i.e., $\zcomd > \frac{1}{2} \lbr \tubel + \tubeconl \rbr$. {In contrast, for partial droplet mobilization, we define $\timemob$ as the time interval required for the center of mass of the mobilized droplet at the downstream to cross the constriction.} Finally, we quantify the mobilization efficacy by measuring $\timemob$ while systematically varying the actuation frequency and amplitude. For droplet mobilization to be most effective, one would want $\timemob$ to be as low as possible, and the droplet to remain intact throughout the mobilization process.

\subsection{Hydrodynamic actuation with oscillatory fluid forcing}

In this subsection, we investigate the impact of oscillatory fluid forcing on droplet mobilization. Here, we study the influence of $\bodyfomgrref$ and $\bodyomegaref$ on $\timemob$. Our simulation results show that $\timemob$ increases monotonically with the forcing frequency $\bodyomegaref$, whereas it decreases monotonically with the forcing magnitude $\bodyfomgrref$.

\subsubsection{Effect of forcing frequency on droplet mobilization}

\begin{figure}[t] 
    \centering
    \includegraphics[width=\linewidth]{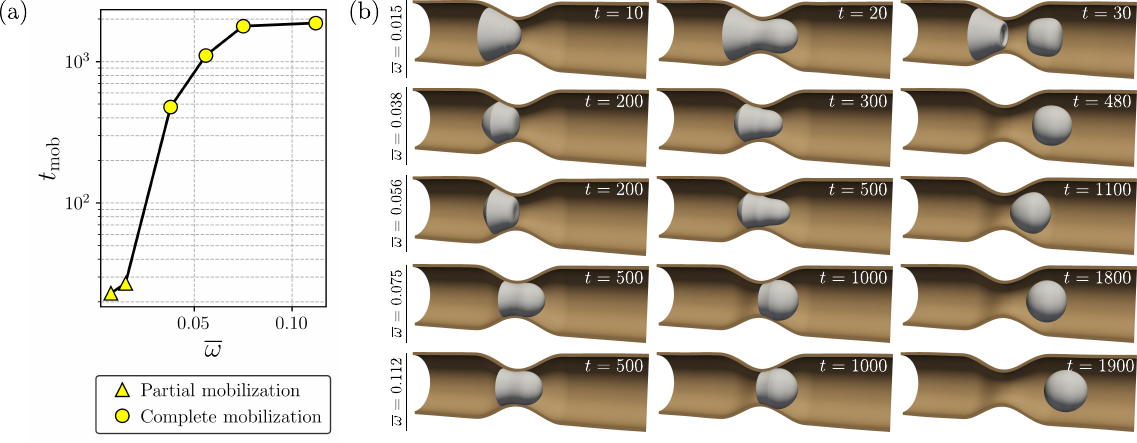}
    \caption{(a) Plot of droplet mobilization time ($\timemob$) in a soft constricted capillary tube versus forcing frequency $\bodyomegaref$, for both partial and complete droplet mobilization regimes. (b) Snapshots of the droplet motion for different values of $\bodyomegaref$. The bubble colored in white is shown by the level set $\phase = 0$ while the tube is shown in brown. The three-dimensional rendering is generated by revolving our two-dimensional axisymmetric simulation result about the symmetry axis. To illustrate the droplet movement, we show a sliced view of the three-dimensional tube.}
    \label{fig:soft_tube_forcingfreq}
\end{figure}

{We first study the effect of $\bodyomegaref$ on $\timemob$ in a soft constricted capillary tube using $\bodyfomgrref = 7.5 \times 10^{-4}$ for all values of $\bodyomegaref$. We perform simulations with six different values of $\bodyomegaref$: $7.5 \times 10^{-3},\, 0.015,\, 0.038,\, 0.056,\, 0.075$ and $0.112$, where $\bodyomegaref = 0.0$ corresponds to no forcing. We plot the droplet mobilization time against $\bodyomegaref$ from our numerical simulations in Fig.~\ref{fig:soft_tube_forcingfreq}(a). Our data show that $\timemob$ increases monotonically with the forcing frequency. This monotonic trend in $\timemob$  against $\bodyomegaref$ has also been reported in rigid constricted tubes \cite{beresnev_grl_2005}. Further, our data in Fig.~\ref{fig:soft_tube_forcingfreq}(a) show that droplet mobilization is partial at low forcing frequencies, i.e., the droplet breaks up and only a fraction of it passes through the tube’s neck; see the top panel in Fig.~\ref{fig:soft_tube_forcingfreq}(b). The mobilization time increases monotonically with $\bodyomegaref$ (see Fig.~\ref{fig:soft_tube_forcingfreq}(a)) because, for each forcing cycle, the time duration when the body force (and, consequently, the inertial forces) is in the direction of pressure drop decreases. Therefore, less net forward motion is imparted to the droplet as $\bodyomegaref$ increases. Our simulation snapshots in Fig.~\ref{fig:soft_tube_forcingfreq}(b) show that the case with $\bodyomegaref=0.015$ undergoes partial mobilization, while the higher forcing frequency cases $\bodyomegaref = 0.056, 0.075$ and $0.112$ exhibit complete droplet mobilization. Further, Fig.~\ref{fig:soft_tube_forcingfreq}(b) shows that, at different times, the droplet-water interface exhibits advancing and receding contact angles that deviate from $\theta$, highlighting the dynamic wetting behavior during droplet mobilization.}

\begin{figure}[t] 
    \centering
    \includegraphics[width=\linewidth]{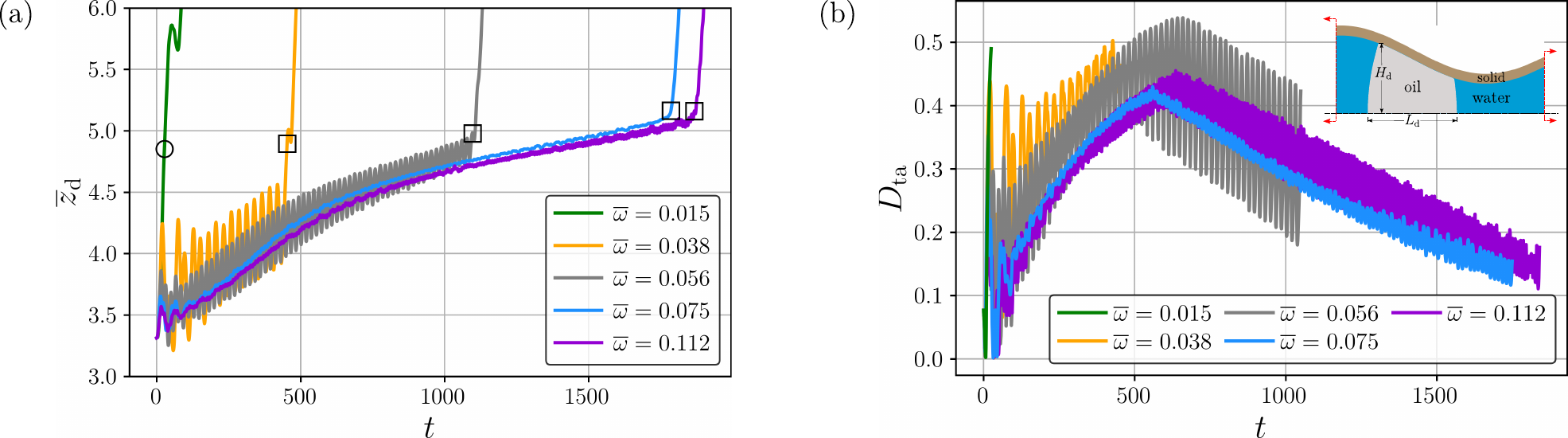}
    \caption{(a) Time evolution of the droplet's axial center of mass $\zcomd$ for different values of forcing frequency $\bodyomegaref$. The black colored circular marker denotes the point of droplet breakup. However, the black colored square marker denotes the point when the droplet's contact separates from the tube walls. (b) Time variation of the primary droplet Taylor deformation parameter $\taylord$. {For $\bodyomegaref = 0.015$ (partial mobilization), $\taylord$ is measured for the primary droplet until its breakup. For the other cases of $\bodyomegaref$ (complete mobilization), however, $\taylord$ is computed until the droplet loses contact with the tube walls.} The inset illustrates the definitions of droplet length ($L_\mathrm{d}$) and height ($H_\mathrm{d}$) used in Taylor deformation parameter ($\taylord$), shown in a two-dimensional axisymmetric view of the tube.}
    \label{fig:soft_tube_forcingfreq_comd}
\end{figure}

We get additional insight into the droplet mobilization process by plotting the time variation of $\zcomd$ in Fig.~\ref{fig:soft_tube_forcingfreq_comd}(a). {We consider the monotonically increasing curve in the $\timemob$--$\bodyomegaref$ plot shown in Fig.~\ref{fig:soft_tube_forcingfreq}(a)}. We have three observations. First, for all cases of $\bodyomegaref$ shown in Fig.~\ref{fig:soft_tube_forcingfreq_comd}(a), $\zcomd$ exhibits oscillations with distinct peaks and troughs, that respectively correspond to the droplet’s advancing motion (when the oscillatory forcing is in the direction of pressure drop) and receding motion (when the oscillatory forcing opposes the pressure drop). Second, as $\bodyomegaref$ increases, the number of oscillatory cycles required to mobilize the droplet increases. This is because the amplitude of the $\zcomd$ peaks and troughs decreases with an increase in $\bodyomegaref$, thereby reducing the net forward motion imparted to the droplet per oscillatory cycle. For example, for $\bodyomegaref = 0.015$, the droplet undergoes breakup and gets partially mobilized before one complete oscillatory cycle of the actuation is over. Third, when the droplet's contact line detaches from the tube walls, the droplet advances significantly over a very short span of time---this point is denoted by square markers in  Fig.~\ref{fig:soft_tube_forcingfreq_comd}(a). {This happens when $\zcomd \approx 5.0$, i.e., the droplet is near the end of the constriction and the capillary resistance offered by the constriction has already vanished}. Consequently, the loss of droplet-wall contact at this point of time causes the release of surface energy, resulting in the abrupt forward motion of the droplet. 

{To understand the influence of the forcing frequency on interfacial distortion during the mobilization process, we plot the droplet's Taylor deformation parameter $\taylord$ in Fig.~\ref{fig:soft_tube_forcingfreq_comd}(b). Here, $\taylord = \frac{L_\mathrm{d} - H_\mathrm{d}}{L_\mathrm{d} + H_\mathrm{d}}$, where $L_\mathrm{d}$ is the maximum droplet's length and $H_\mathrm{d}$ is the height of the droplet from the symmetry axis, respectively; see the inset in Fig.~\ref{fig:soft_tube_forcingfreq_comd}(b). When $\bodyomegaref = 0.015$, $\taylord$ exhibits a large peak-to-trough variation within an oscillation cycle while the droplet is in the neck of the constriction. The longer forcing period provides sufficient time for the droplet to deform, ultimately leading to breakup. As $\bodyomegaref$ increases, the forcing time period decreases and the peak-to-trough variation of $\taylord$ in each oscillation cycle decreases---this indicates that the time available for droplet deformation becomes limited, thereby inhibiting droplet breakup. Further, the moving-average plot of $\taylord$ after filtering out high-frequency oscillations becomes non-monotonic at higher forcing frequencies, i.e., when $\bodyomegaref = 0.056,\, 0.075$ and $0.112$. In these higher forcing frequency cases, the peak $\taylord$ value is attained when the droplet is still confined within the neck of the constriction. Subsequently, $\taylord$ decreases temporally as the droplet fully enters the diverging section of the tube, where $H_\mathrm{d}$ progressively increases.}

\subsubsection{Effect of forcing amplitude on droplet mobilization}

\begin{figure}[t] 
    \centering
    \includegraphics[width=\linewidth]{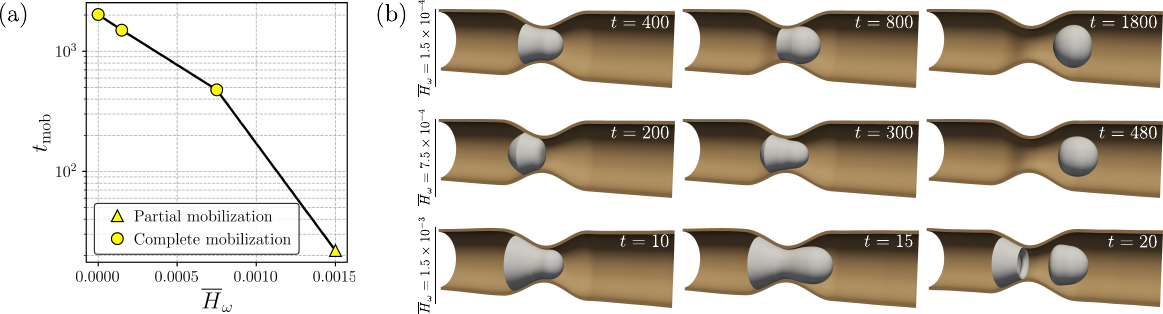}
    \caption{(a) Plot of droplet mobilization time ($\timemob$) in a soft constricted capillary tube versus forcing amplitude $\bodyfomgrref$, for both partial and complete droplet mobilization regimes. (b) Snapshots of the droplet's motion for different values of $\bodyfomgrref$. The bubble colored in white is shown by the level set $\phase = 0$ while the soft tube is colored in brown. The three-dimensional rendering is generated by revolving our two-dimensional axisymmetric simulation result about the symmetry axis. To illustrate the droplet movement, we show the sliced view of the three-dimensional tube.}
    \label{fig:soft_tube_forcingamp}
\end{figure}

\begin{figure}[t] 
    \centering
    \includegraphics[width=\linewidth]{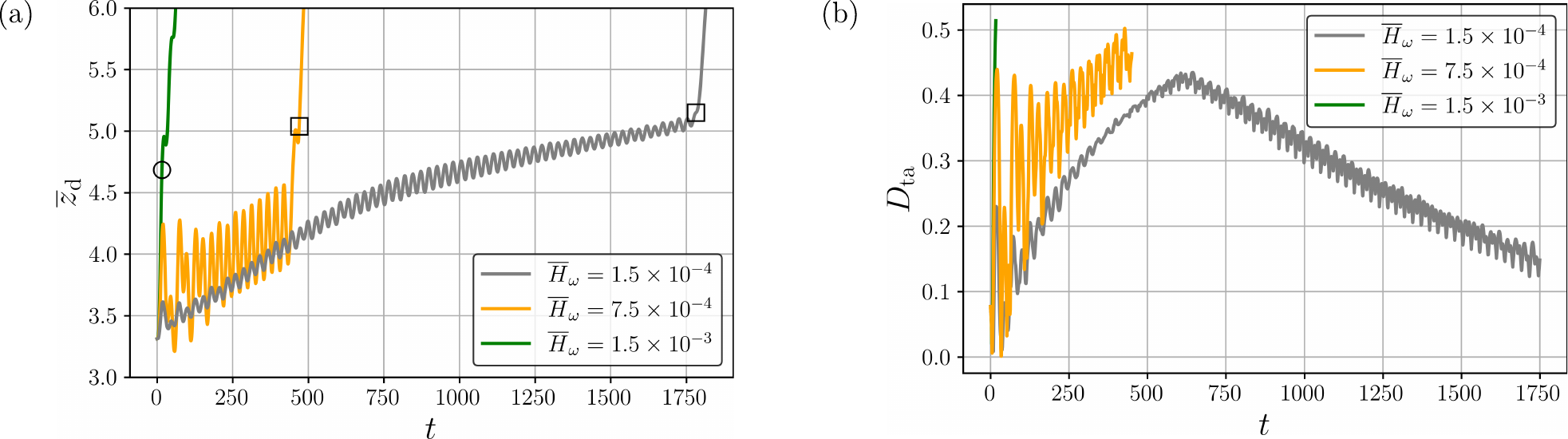}
    \caption{(a) Time evolution of the droplet's axial center of mass $\zcomd$ for different values of forcing amplitudes $\bodyfomgrref$. The black colored circular marker denotes the point of droplet breakup. However, the black colored square marker denotes the point when the droplet's contact separates from the tube walls. (b) Time variation of primary droplet Taylor deformation parameter $\bodyfomgrref$. {For $\bodyomegaref = 1.5 \times 10^{-3}$ (partial mobilization), $\taylord$ is measured for the primary droplet until its breakup. For $\bodyomegaref = 1.5 \times 10^{-4}$ and $7.5 \times 10^{-4}$ (complete mobilization), however, $\taylord$ is computed until the droplet loses contact with the tube walls.}}
    \label{fig:soft_tube_forcingamp_comd}
\end{figure}

Here, we study the effect of $\bodyfomgrref$ on $\timemob$ for a fixed value of $\bodyomegaref = 0.038$. We perform simulations with three different values of $\bodyfomgrref$: $1.5 \times 10^{-4},\, 7.5 \times 10^{-4}$ and $1.5 \times 10^{-3}$. The results from our numerical simulations are shown in Fig.~\ref{fig:soft_tube_forcingamp}(a). The data indicate that the times for droplet mobilization decrease with an increase in $\bodyfomgrref$. {However, for the largest value of $\bodyfomgrref$ the mobilization is only partial.} 
Droplet mobilization time decreases with $\bodyfomgrref$ because a higher value of $\bodyfomgrref$ increases the net forward displacement of the droplet per oscillating cycle. 
We next plot the time evolution of $\zcomd$ in Fig.~\ref{fig:soft_tube_forcingamp_comd}(a). Our data show that an increase in $\bodyfomgrref$ decreases the number of oscillating cycles required for droplet mobilization, but increases the amplitude of both advancing and receding motions of the droplet. When $\bodyfomgrref = 1.5 \times 10^{-3}$, the droplet undergoes breakup even before one complete forcing cycle is over. To better understand the effect of forcing amplitude on droplet deformation, we plot the time evolution of $\taylord$ in Fig.~\ref{fig:soft_tube_forcingamp_comd}(b). {When $\bodyfomgrref = 1.5 \times 10^{-4}$, the peak-to-trough variation of $\taylord$ during an oscillation cycle is small and the moving average plot of $\taylord$, after filtering out the high-frequency oscillations, exhibits a non-monotonic temporal variation. The forcing in this case is too weak to cause droplet breakup. When the forcing amplitude is increased to $\bodyfomgrref = 7.5 \times 10^{-4}$, the peak-to-trough variation of $\taylord$ during an oscillation cycle is relatively higher than at $\bodyfomgrref = 1.5 \times 10^{-4}$, and the moving average of $\taylord$ increases monotonically with time. The absence of non-monotonicity in the moving average of $\taylord$ over time, indicates that the droplet's contact separates from the tube walls only after it has just passed the neck of the constriction.} Although the droplet deforms appreciably in this case, the forcing is still insufficient to cause droplet breakup. For the highest value of forcing amplitude, i.e., when $\bodyfomgrref = 1.5 \times 10^{-3}$, $\taylord$ rises very steeply in a monotonic fashion due to the rapid increase in $L_\mathrm{d}$. The forcing in this case, due to the very large value of $\bodyfomgrref$, is sufficient to cause the droplet's breakup. Our results in Fig.~\ref{fig:soft_tube_forcingamp_comd}(b) are consistent with previous observations in rigid constrictions \cite{al_etal_jpse_2021}, where it has been shown that increasing the forcing amplitude increases the droplet’s Weber number---defined as the ratio of inertial forces to the surface tension forces---thereby enhancing the possibility of droplet breakup.

\subsection{Dynamic wall actuation through an oscillatory follower-load traction}

In this subsection, we study the effect of dynamic wall actuation on droplet mobilization by varying $\presfolomgref$ and $\wallomgref$. We impose the traction as a follower-load over an upstream segment of the capillary tube (see Eq.~\eqref{eqn:followerloadpres} and Fig.~\ref{fig:schematic}). We selected the length of this actuation zone based on preliminary simulations, which showed that this segment length is an efficient and simple way to mobilize the droplet. In all our simulations, $\presfolomg$ is at least two times larger than $p_-$. Our chosen value of $\presfolomg$ ensures that the droplet mobilization dynamics are dominated by the actuation rather than the pressure drop. Unlike hydrodynamic actuation, dynamic wall actuation cannot be applied to a rigid tube. In what follows, our simulations show that $\timemob$ decreases monotonically with an increase in $\presfolomgref$. However, $\timemob$ exhibits a non-monotonic dependence on $\wallomgref$, highlighting a resonance-like behavior when $\hat{\Omega}$ is close to $\hat{\Omega}_\text{res}$.

\subsubsection{Effect of traction frequency ratio on droplet mobilization}

\begin{figure} 
    \centering
    \includegraphics[width=\linewidth]{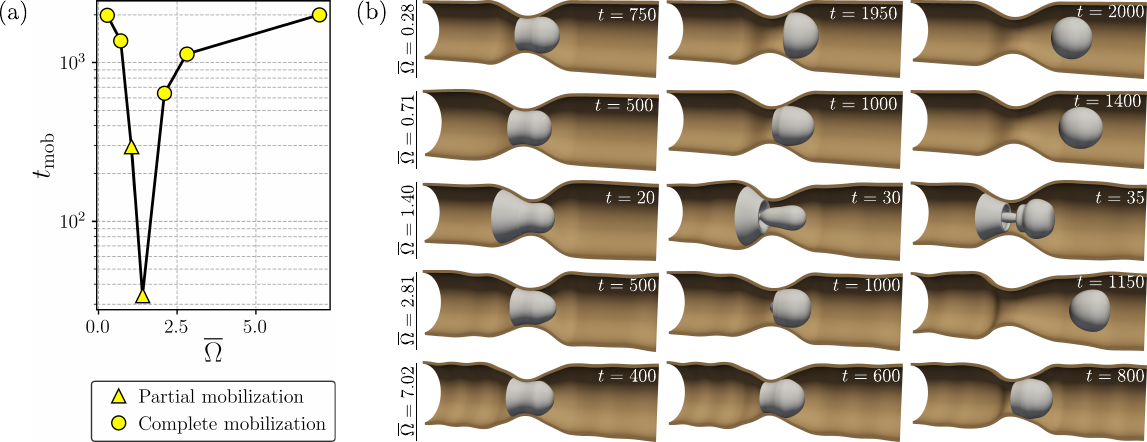}
    \caption{(a) Plot of droplet mobilization time ($\timemob$) in a soft constricted capillary tube versus traction frequency ratio $\wallomgref$, for both partial and complete droplet mobilization regimes. (b) Snapshots of the droplet's motion for different values of $\wallomgref$. The bubble colored in white is shown by the level set $\phase = 0$ while the tube is shown in brown. The three-dimensional rendering is generated by revolving our two-dimensional axisymmetric simulation result about the symmetry axis. To illustrate the droplet movement, we only show the sliced view of the three-dimensional tube.}
    \label{fig:soft_tube_dynact_freq}
\end{figure}

We first study the effect of $\wallomgref$ on $\timemob$ using a fixed value of $\presfolomgref = 0.11$ for all values of $\wallomgref$. We conduct simulations for seven different values of $\wallomgref$: 0.28, 0.71, 1.05, 1.40, 2.10, 2.81 and 7.02. The cyclic wall motion modulates the pressure field and facilitates droplet mobilization through the constriction. Our simulation data in Fig.~\ref{fig:soft_tube_dynact_freq}(a). 
show a non-monotonic dependence of $\timemob$ on $\wallomgref$: $\timemob$ decreases as $\wallomgref$ is increased from $0.28$ to $1.40$, then increases again as $\wallomgref$ is raised beyond $1.40$. {Further, the cases of $\wallomgref = 1.05$ and $1.40$ show partial droplet mobilization, while all other cases exhibit complete droplet mobilization}. When we actuate the tube with $\wallomgref = 1.40$, the tube’s deformation is greatly amplified, the droplet undergoes breakup and subsequently mobilizes the fastest; see the middle panel in Fig.~\ref{fig:soft_tube_dynact_freq}(b). We term $\wallomgref = 1.40$ as the coupled droplet-tube’s resonant frequency, which we estimate by tracking $\timemob$ and $\zcomd$ from our simulation data.

\begin{figure}[t] 
    \centering
    \includegraphics[width=\linewidth]{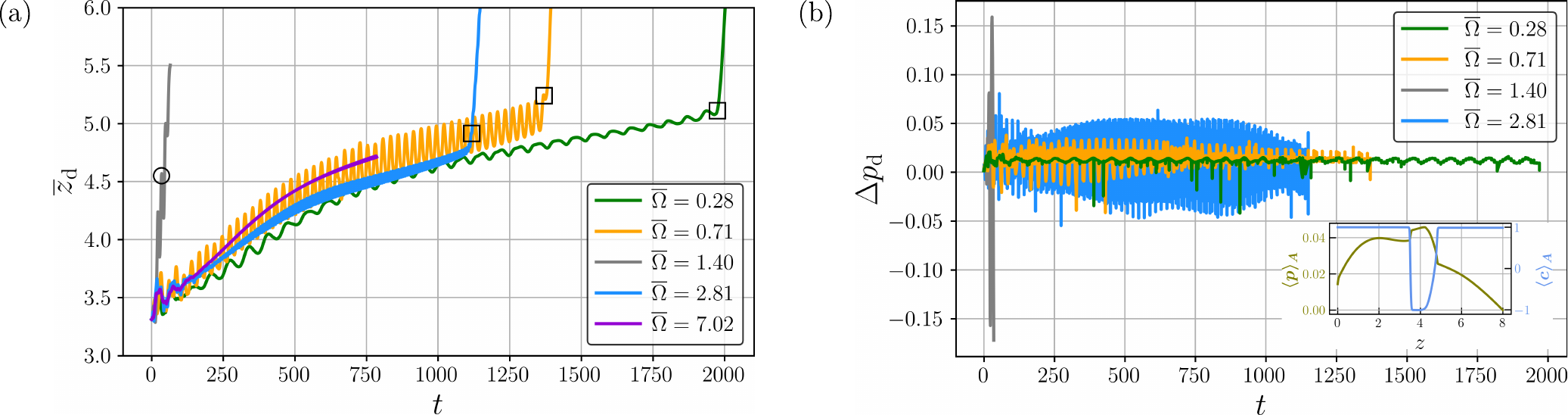}
    \caption{Panel (a) shows the time evolution of the droplet's axial center of mass $\zcomd$ for different values of traction frequency ratio $\wallomgref$. The black colored circular marker denotes the point of droplet breakup. However, the black colored square marker denotes the point when the droplet's contact separates from the tube walls. Panel (b) shows the time evolution of the area-averaged pressure drop across the droplet for different values of $\wallomgref$. The inset shows the axial variation of area-averaged pressure $\langle p \rangle_A$ and area-averaged phase-field $\langle \phase \rangle_A$ for $\wallomgref = 0.71$ at $t = 500$.}
    \label{fig:soft_tube_dynact_freq_comd}
\end{figure}

{The dependence of $\timemob$ with $\wallomgref$ can be quantitatively understood from the temporal variation of $\zcomd$ in Fig.~\ref{fig:soft_tube_dynact_freq_comd}(a). At the resonant frequency, the droplet undergoes break up within the first three oscillation cycles, and subsequently exhibits rapid axial advancement. However, the off-resonant actuation cases require many more oscillatory cycles to achieve droplet mobilization. Fig.~\ref{fig:soft_tube_dynact_freq_comd}(a) further shows that the amplitude of each peak and trough in an oscillation is low at high off-resonant frequencies, i.e., $\wallomgref = 2.81$ and $\wallomgref = 7.02$, and high at low off-resonant frequencies, i.e., $\wallomgref = 0.28$ and $\wallomgref = 0.71$. To get additional insight into how dynamic wall actuation controls the droplet mobilization time, we plot the time variation of the area-averaged pressure drop across the droplet $\Delta p_\text{d}$ for different values of $\wallomgref$; see Fig.~\ref{fig:soft_tube_dynact_freq_comd}(b). We define $\Delta p_\text{d} = \langle p \rangle_A \lbr \xz_\text{d,r}\rbr - \langle p \rangle_A \lbr \xz_\text{d,a}\rbr$, where $\langle p \rangle_A \lbr \xz \rbr = \frac{\int_0^{R_0^{'} \lbr \xz \rbr} p \lbr \xr, \xz \rbr 2 \pi \xr \mathrm{d}\xr}{\int_0^{R_0^{'} \lbr \xz \rbr} 2 \pi \xr \mathrm{d}\xr}$ is the area-averaged pressure at $\xz$, $R_0^{'}$ is the radius of the deformed tube, and $\xz_\text{d,r}$, $\xz_\text{d,a}$ denote the receding and advancing contact line positions of the droplet. We compute these contact line positions from the area-averaged phase-field $\langle \phase \rangle_A$: $\xz_\text{d,r}$ is located where $\langle \phase \rangle_A = 0$ and $\frac{\mathrm{d} \langle \phase \rangle_A}{\mathrm{d} \xz} < 0$, whereas $\xz_\text{d,a}$ is located where $\langle \phase \rangle_A = 0$ and $\frac{\mathrm{d} \langle \phase \rangle_A}{\mathrm{d} \xz} > 0$; see the inset in Fig.~\ref{fig:soft_tube_dynact_freq_comd}(b). A positive value of $\Delta p_\text{d}$ in Fig.~\ref{fig:soft_tube_dynact_freq_comd}(b) corresponds to the droplet's advancing motion, while a negative value corresponds to the droplet's receding motion. Here, $\Delta p_\text{d}$ is a measure of the droplet’s driving force. As shown in Fig.~\ref{fig:soft_tube_dynact_freq_comd}(b), the resonant case with $\wallomgref = 1.40$ exhibits the largest pressure peaks, indicating the strongest driving force, which explains the most pronounced droplet deformation and the fastest mobilization. We compute the root-mean-square of pressure drop over the mobilization time interval as $\Delta p_\mathrm{d,rms} = \sqrt{\frac{1}{\timemob} \int_0^{\timemob} \lbr \Delta p_\text{d} \rbr^2 \mathrm{d}t}$. From our simulations, $\Delta p_\text{d,rms} = 0.012$ for $\wallomgref = 0.28$, $\Delta p_\text{d,rms} = 0.016$ for $\wallomgref = 0.71$, $\Delta p_\text{d,rms} = 0.075$ for $\wallomgref = 1.40$ and $\Delta p_\text{d,rms} = 0.029$ for $\wallomgref = 2.81$. Our data reveals an inverse relationship between $\Delta p_\mathrm{d,rms}$ and $\timemob$, which explains the non-monotonic dependence of $\timemob$ on $\wallomgref$.}

\subsubsection{Effect of traction amplitude ratio on droplet mobilization}

\begin{figure}[t] 
    \centering
    \includegraphics[width=\linewidth]{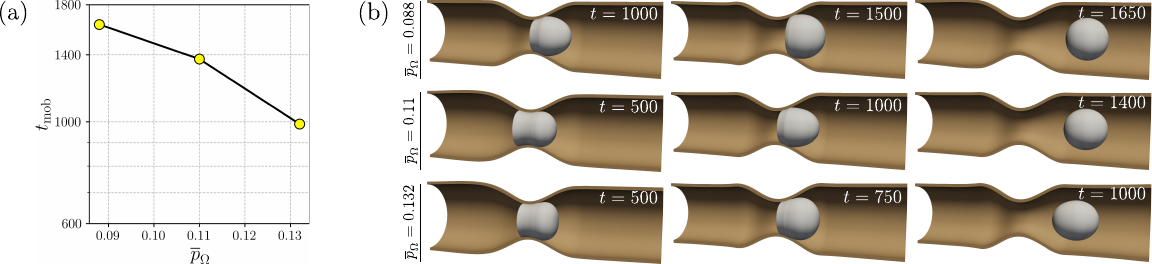}
    \caption{(a) Plot of droplet mobilization time ($\timemob$) in a soft constricted capillary tube versus traction amplitude ratio $\presfolomgref$, for complete droplet mobilization regime. (b) Snapshots of the droplet's motion for different values of $\presfolomgref$. The bubble colored in white is shown by the level set $\phase = 0$ while the tube is shown in brown. The three-dimensional rendering is generated by revolving our two-dimensional axisymmetric simulation result about the symmetry axis. To illustrate the droplet movement, we only show the sliced view of the three-dimensional tube.}
    \label{fig:soft_tube_dynact_amp}
\end{figure}

\begin{figure}[t] 
    \centering
    \includegraphics[width=0.5\linewidth]{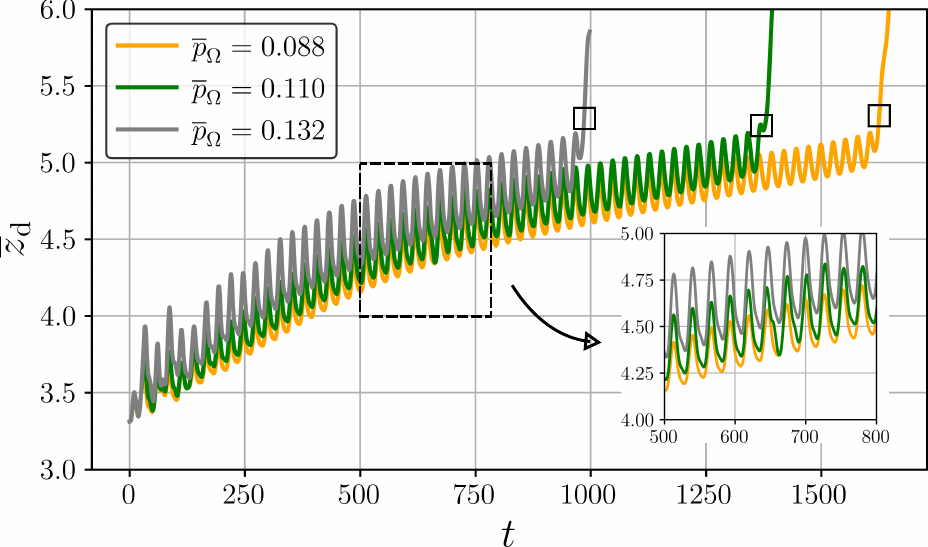}
    \caption{Time evolution of the droplet's axial center of mass $\zcomd$ for different values of traction amplitude ratio $\presfolomgref$. The black colored square marker denotes the point when the droplet's contact separates from the tube walls.}
    \label{fig:soft_tube_dynact_amp_comd}
\end{figure}

To study the effect of $\presfolomgref$ on $\timemob$, we perform simulations with three different values of $\presfolomgref$: $0.088,\, 0.11$ and $0.132$ for a fixed value of $\wallomgref = 0.71$. The maximum traction amplitude we can use here is constrained by the largest solid deformation that can be supported by our boundary-fitted fluid–structure algorithm without affecting the mesh quality. The mobilization times are shown in Fig.~\ref{fig:soft_tube_dynact_amp}(a). Additionally, our simulation snapshots in Fig.~\ref{fig:soft_tube_dynact_amp}(b) show that the droplet in the case of the lowest traction amplitude ratio leaves the constriction last, while the droplet in the case of highest traction amplitude ratio leaves the constriction first. Our simulations report two findings that are similar to what we have previously found in the case of hydrodynamic actuation. First, we observe a monotonic decrease of $\timemob$ with an increase in $\presfolomgref$. This is because, as $\presfolomgref$ increases, the net forward displacement of the droplet per oscillating cycle increases. We confirm this observation from the variation of peak-trough amplitude of $\zcomd$, which we show in Fig.~\ref{fig:soft_tube_dynact_amp_comd}. Second, an increase in $\presfolomgref$ reduces the total number of oscillating cycles required for droplet mobilization. Unlike hydrodynamic actuation, a change in $\presfolomgref$ in dynamic wall actuation, does not alter the mobilization regime: for the off-resonant frequency we consider here, we observe complete droplet mobilization across all tested values of $\presfolomgref$.

\subsubsection{Phase diagram of droplet mobilization}

\begin{figure}[t] 
    \centering
    \includegraphics[width=0.9\linewidth]{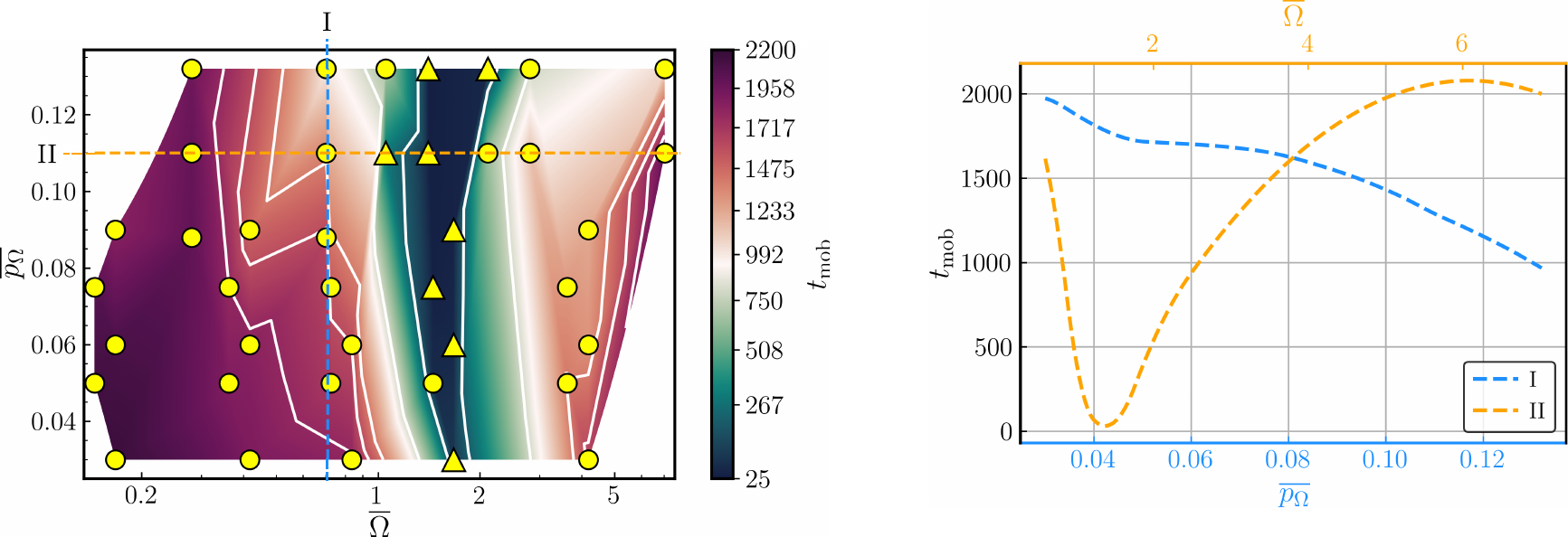}
    \caption{Control of droplet mobilization in a dynamic wall actuation mechanism. (Left) Dependence on two dimensionless parameters: traction amplitude ratio $\presfolomgref$ and traction frequency ratio $\wallomgref$ on the droplet mobilization time $\timemob$. Each yellow marker in the plot indicates a numerical solution for a $\presfolomgref$-$\wallomgref$ pair. The circular markers denote complete droplet mobilization regime, while triangular markers denote partial droplet mobilization regime. (Right) Variation of $\timemob$ with respect to $\presfolomgref$ (in blue) and $\wallomgref$ (in orange) along two cut-lines I and II, made when $\wallomgref = 0.7$ and $\presfolomgref = 0.11$, respectively.}
    \label{fig:soft_tube_dynact_phdiagram}
\end{figure}

We construct a phase diagram on the $\lbr \presfolomgref, \wallomgref \rbr$ space and mark the corresponding droplet mobilization regimes in Fig.~\ref{fig:soft_tube_dynact_phdiagram}. This phase diagram is a map for selecting the traction frequency and traction amplitude ratios to yield a desired droplet mobilization time. We show the results of our $45$ high-resolution simulations (represented by circular or triangular markers in the plot), each corresponding to a $\presfolomgref$-$\wallomgref$ pair. We identify a narrow vertical band colored in dark green ($\timemob < 50$) which denotes fast droplet mobilization accompanied by droplet breakup. We term this region as a resonant band, because the resonant frequency at which the droplet mobilizes the fastest lies in this band. Additionally, we identify two control principles from this phase diagram. First, fastest mobilization is achieved by actuating the tube with the largest traction amplitude and a resonant frequency. Second, slowest mobilization is achieved by actuating the tube with the smallest traction amplitude and a frequency that is farthest away from resonance. As shown by the line plots in Fig.~\ref{fig:soft_tube_dynact_phdiagram}, the trend in $\timemob$ is consistent with our earlier observations: $\timemob$ monotonically decreases with an increase in $\presfolomgref$ (line-plot I), while $\timemob$ is non-monotonic in $\wallomgref$ due to the resonance effect (line-plot II). The mobilization time has a stronger dependence on $\wallomgref$ than on $\presfolomgref$: in line-plot I, as $\wallomgref$ is moved away from the resonant frequency, $\timemob$ changes, roughly, by an order of magnitude; in line plot II, as $\presfolomgref$ is decreased, the corresponding decrease in $\timemob$ is relatively weaker.  Overall, our data show that dynamic wall actuation permits us to control both droplet mobilization times and mobilization regimes for droplet microfluidics applications.

\section{Conclusions}

In this study, we investigate the influence of two actuation mechanisms: hydrodynamic actuation and dynamic wall actuation, on droplet mobilization in a deformable, constricted capillary tube. Here, hydrodynamic actuation is driven by an oscillatory fluid force, whereas dynamic wall actuation is driven by an oscillatory follower-load traction applied along the tube walls. Using high-resolution numerical simulations, we examine how actuation amplitude and frequency affect the time required  to mobilize the droplet beyond the constriction. For the simulation parameters considered in this study, our data show that, in the case of hydrodynamic actuation, droplet mobilization time increases monotonically with the forcing frequency and decreases monotonically with the forcing amplitude. In dynamic wall actuation, the mobilization time decreases monotonically with the traction amplitude. However, we observe a non-monotonic dependence of mobilization time on traction frequency due to a resonance phenomenon; when the traction frequency approaches the coupled fluid-tube’s natural frequency, the mobilization time decreases, reaching a minimum at resonance, and then increases as the frequency moves away from resonance. Our simulations also capture two mobilization regimes during the mobilization process: (a) partial droplet mobilization, where the droplet breaks up before it crosses the constriction's neck, and (b) complete droplet mobilization, where the droplet remains intact. In hydrodynamic actuation, the transition between these mobilization regimes is controlled by varying either forcing frequency or forcing amplitude. In dynamic wall actuation, however, the traction amplitude does not alter the mobilization regime, but it is the traction frequency which does; actuating the tube with an off-resonant frequency results in complete droplet mobilization; actuating the tube with a frequency in the resonant band results in partial droplet mobilization. In bio-microfluidic and lab-on-chip applications, partial mobilization is often undesirable,  because the mobilization process leaves behind a thin oil film adhered to the tube walls. To achieve complete droplet mobilization in such cases, one strategy is to dynamically introduce additional droplets in the system that coalesce with and displace this residual film.

Although we have assumed a constant actuation amplitude in this study, future actuation mechanisms could use a spatially varying actuation amplitude that could adaptively modulate the local tube deformation---similar to a feedback-controlled actuator network. Such adaptive strategies are used by biological systems for removing fluid boluses \cite{lei_etal_jnm_2022}. Further, while we have employed a sinusoidal waveform for the oscillatory actuation, alternative waveforms—such as asymmetric pulses, sawtooth, square waves, or combinations thereof—may enhance droplet mobilization and warrant future exploration.

Our study is one of the first investigations of droplet transport in dynamically actuated deformable constrictions. We identify two key directions for future research: (1) droplet transport in interconnected networks of deformable constrictions, and (2) droplet mobilization using surface acoustic wave (SAW) actuation. Regarding the first topic, networks of deformable constrictions are commonly found in pumpless microfluidic chips \cite{choi_etal_adhm_2017}, soft porous media, plant biological systems \cite{keiser_etal_jfm_2022}, compliant vascular structures \cite{li_etal_pnas_2021} and chemical engineering processes—applications in which droplet and bubble transport must be better understood. In these applications, many polydisperse droplets are typically present, and identifying an optimal strategy for mobilizing all these droplets is essential. With reference to our second topic, our current model assumes moderate actuation frequencies, where the effects of fluid compressibility, acoustic streaming, and thermo-viscous heating are considered to be negligible. However, at higher frequencies typical of SAW devices, these effects may become important. To capture the resulting flow physics with these effects, our model must then be extended to include: (a) fluid compressibility, phase-change dynamics and thermal effects, which can be addressed using a thermally-coupled Navier–Stokes–Korteweg model \cite{bueno_etal_compmech_2015, liu_etal_cmame_2015}; and (b) coupled solid–fluid acoustodynamics to resolve acoustic wave propagation in both the solid and fluid domains \cite{howe_book_1998}. Incorporating these multiphysics interactions will then enable the predictive design of SAW-driven droplet manipulation platforms \cite{ding_etal_sawrev_2013} for applications such as biochemical assays \cite{huang_etal_2021}, emulsion synthesis \cite{taha_etal_rev_2020}, and lab-on-chip technologies \cite{weser_etal_ultrasonics_2020, wang_etal_labonchip_2011}.

\vspace{0.05\textwidth}

\noindent \textbf{Acknowledgements} \\
This research was supported by the National Science Foundation (Award no. CBET 2012242). S.R.B was supported by the Bilsland Dissertation Fellowship from the Office of Interdisciplinary Graduate Programs at Purdue University. This work used Bridges-2 system at the Pittsburgh Supercomputing Center (PSC) through allocation no. MCH220014 from the Advanced Cyberinfrastructure Coordination Ecosystem: Services $\&$ Support (ACCESS) program \cite{boerner_access_2023}, which is supported by National Science Foundation grant nos. 2138259, 2138286, 2138307, 2137603, and 2138296. The opinions, findings, and conclusions, or recommendations expressed are those of the authors and do not necessarily reflect the views of the National Science Foundation. \\

\noindent \textbf{Author Contributions} \\
\textbf{Sthavishtha R. Bhopalam}: conceptualization (lead); data curation (lead); formal analysis (lead); investigation (lead); methodology (equal); software (lead); visualization (lead); writing - original draft (lead); writing - review \& editing (equal). \textbf{Ruben Juanes}: conceptualization (lead); investigation (supporting); supervision (supporting); writing - original draft (supporting); writing - review \& editing (equal). \textbf{Hector Gomez}: conceptualization (lead); formal analysis (supporting); funding acquisition (lead); investigation (supporting); methodology (equal); project administration (lead); resources (lead); supervision (lead); writing - original draft (supporting); writing - review \& editing (equal). All authors reviewed and approved the final version of the manuscript. \\

\noindent \textbf{Competing Interests} \\ 
The authors declare that they have no known competing financial interests or personal relationships that could have appeared to influence the work reported in this paper. \\

\noindent \textbf{Data availability statement} \\
Data for this article is available at \href{https://doi.org/10.4231/JMDG-P694}{https://doi.org/10.4231/JMDG-P694}.

\bibliography{ref}
\bibliographystyle{apsrev4-2}

\appendix

\section{Pressure scaling in dynamic wall actuation mechanism}
\label{sec:appendix_a}

In this section, we justify the choice of our pressure scale in non-dimensionalizing the traction pressure. We consider a straight axisymmetric annular tube of inner and outer radii, $\hat R_m$ and $\hat R_o = \hat R_m + \hat \tubeh$ respectively. This annular tube is subjected to an internal and external pressure of $\hat p_i$ and $\hat p_\Omega$, respectively. Assuming the tube's material to be described by a homogeneous, isotropic, linear and elastic solid, the classical Lame' solution for the tube's radial displacement \cite{hetnarski_book_2016} is given by 
\begin{equation}
  \hat u_R(\hat r)=
  \frac{1-\nu}{\hat E}\,
  \frac{\hat p_i \hat R_m^{2}-\hat p_\Omega \hat R_o^{2}}
       {\hat R_o^{2}-\hat R_m^{2}}\, \hat r
  +\,
  \frac{1+\nu}{\hat E}\,
  \frac{(\hat p_i-\hat p_\Omega)\hat R_m^{2}\hat R_o^{2}}
       {(\hat R_o^{2}-\hat R_m^{2})\,\hat r}.
    \label{eqn:lame}
\end{equation}

We make two additional assumptions: zero internal pressure on the tube, i.e., $\hat p_i=0$ and thin-wall limit of the tube, i.e., $\hat h\ll\hat R_o$. With these assumptions, Eq.~\eqref{eqn:lame} now simplifies to
\begin{equation}
  \hat u_R(\hat r) \approx
  -\frac{\hat p_\Omega\,\hat R_o}
        {2\hat E}
  \lbr \lbr 1 - \nu \rbr \frac{\hat r}{\hat \tubeh} + \lbr 1 + \nu \rbr \frac{\hat R_o^2}{\hat r \,\hat \tubeh}\rbr.
    \label{eqn:lame_simplify}
\end{equation}
where we have used $(\hat R_o^{2}-\hat R_m^{2}) \approx 2 \hat R_o \, \hat \tubeh$. We now compute the radial displacement of the tube on the outer surface $u_{R,o} = u_R(\hat R_o)$ from Eq.~\eqref{eqn:lame_simplify} and subsequently simplify to get,
\begin{equation}
  \hat u_{R,o} \approx
  -\frac{\hat p_\Omega\,\hat R_o^2}
        {\hat E \hat \tubeh}.
    \label{eqn:lame_simplify2}
\end{equation}

Eq.~\eqref{eqn:lame_simplify2} gives us the scaling relation
\begin{equation}
  {\;
    \frac{\hat u_{R,o}}{\hat R_o}
    \;\sim\;
    \frac{\hat p_\Omega\,(\hat R_m + \hat h)}
         {\hat E\,\hat h}
  \;}
\end{equation}

Choosing the reference pressure
\[
  \hat p_\mathrm{ref} = \frac{\hat E\,\hat h}{\hat R_m+\hat h},
\]
yields us the dimensionless parameter
\begin{equation}
  {\;
    \presfolomgref
    = \hat p_\Omega p_\mathrm{ref}^{-1}
  \;}
  \ \ \text{or equivalently} \ \
  {\;
    \presfolomgref
    = \frac{\hat p_\Omega \lbr \hat R_m+\hat h \rbr}{\hat E\,\hat h}
  \;}
\end{equation}
termed as the \emph{traction amplitude ratio}. This dimensionless parameter measures the ratio of the applied traction to the elastic restoring stress of the tube wall.

\end{document}